# THE $z = 0.8596$ DAMPED LYMAN ALPHA ABSORBING GALAXY TOWARD PKS 0454+039[1]

Charles C. Steidel[2,3,4]

MIT, Physics Department, Room 6-201, Cambridge, MA 02139

David V. Bowen, J. Chris Blades, and Mark Dickinson [2]

Space Telescope Science Institute, 3700 San Martin Drive, Baltimore, MD 21218

## ABSTRACT

We present *Hubble Space Telescope* and ground–based data on the $z_{abs} = 0.8596$ metal line absorption system along the line of sight to PKS 0454+0356. The system is a moderate redshift damped Lyman alpha system, with N(HI) = $(5.7 \pm 0.3) \times 10^{20}$ cm$^{-2}$ as measured from the *Faint Object Spectrograph* spectrum. We also present ground–based images which we use to identify the galaxy which most probably gives rise to the damped system; the most likely candidate is relatively underluminous by QSO absorber standards ($M_B \sim -19.0$ for $q_0 = 0.5$ and $H_0 = 50$ km s$^{-1}$Mpc$^{-1}$), and lies $\sim 8.5h^{-1}$ kpc in projection from the QSO sightline. Ground–based measurements of Zn II, Cr II, and Fe II absorption lines from this system allow us to infer abundances of [Zn/H]=$-1.1$, [Cr/H]=$-1.2$, and [Fe/H]=$-1.2$, indicating overall metallicity similar to damped systems at $z > 2$, and that the depletion of Cr and Fe onto dust grains may be even *less* important than in many of the high redshift systems of comparable metallicity. Limits previously placed on the 21-cm optical depth in the $z = 0.8596$ system, together with our new N(H I) measurement, suggest a very high spin temperature for the H I, $T_S > 580$ K.

*Subject headings:* galaxies: evolution–galaxies: distances and redshifts–quasars: absorption lines

## 1. Introduction

Recent workers (e.g., Pettini *et al.* 1994, Pettini *et al.* 1990, Meyer and Roth 1990) have made substantial progress toward gaining an understanding of the chemical evolution of normal galaxies by establishing that, at $z \sim 2$, the overall abundances in damped Lyman $\alpha$ (DLA) systems, which are generally envisioned to be the progenitors of disk galaxies of the present epoch (e.g., Wolfe 1993), spanned a wide range of values, but had an "average" metallicity of $\sim 1/10$ of the solar values. They have also shown that the depletion of refractory elements is much less severe than in the interstellar medium (ISM) of the Milky Way Galaxy, indicating that, while dust does appear to be present, the dust/gas ratio is $\sim 1/10$ of the local value. A similar result has been obtained based on consideration of the statistical reddening of background QSOs having DLAs along their line of sight (Pei *et al.* 1993). The metallicity and dust content determinations are based on measurements of the weak lines of Zn II $\lambda\lambda 2026, 2062$ and Cr II $\lambda\lambda\lambda 2056, 2062, 2066$ in the DLA systems; Zn is particularly useful as a "tracer" of Fe–peak elements because it is generally not incorporated into dust grains, so that the gas–phase abundance is very nearly the total abundance, while Cr is heavily depleted in the ISM ($\sim 99\%$ in the solid form) but is found to track the Zn abundance in stars. Thus, as discussed by Pettini *et al.* 1994, measurement of just these 2 species in combination with a measurement of N(H I) for the DLA system simultaneously allows one to infer both the extent to which chemical evolution has progressed, and the extent to which dust depletion is important, in the ISM of high redshift galaxies.

Clearly, the next step is to use the same techniques to investigate the evolutionary status of galaxies at both lower and higher redshift, so that a pattern of chemical enrichment in disk galaxies over time can be established. While this can be pursued to high redshift in a relatively straightforward manner (and the samples of high redshift DLAs are becoming larger), progress at lower $z$ was hindered by the fact that until *Hubble Space Telescope* (HST), the measurement of accurate H I column densities was extremely difficult, as was the detection of the Zn II and Cr II absorption lines at moderate redshift ($0.6 \leq z \leq 1$). The only example of a moderate redshift DLA system to be investigated in detail was 3C 286 (Meyer & York 1992), for which very low abundances ([Zn/H]$= -1.2$) and dust content were found in the system with $z_{abs} = 0.692$. As pointed out by these authors, the values found were very close to what might be expected for a typical DLA system at $z \sim 2$, but are much lower than would be expected if the galaxy producing the absorption has a star formation history similar to the Milky Way. Later, from high–resolution ground–based imaging, Steidel *et al.* (1994) found that the 3C 286 DLA system is produced by a relatively luminous galaxy of very low surface brightness, possibly explaining the low chemical abundances. It is only by obtaining data for more examples of moderate redshift DLA systems that one can be sure whether or not the properties of the 3C 286 system are anomalous.

In this paper, we present observations of a new DLA system, at $z = 0.8596$, along the line of sight to PKS 0454+0356. The system was flagged as an almost certain DLA when Steidel *et al.* (1993) detected Zn II and Cr II lines associated with the previously known (Weymann *et al.* 1978, Steidel & Sargent 1992) metal line system while searching for weak Ca II K and H absorption due to a $z = 0.07$ dwarf galaxy discovered very near the QSO sightline. We have now obtained the *HST* data necessary to determine the chemical abundances and dust content in the $z_{abs} = 0.8596$ system, and we present new, deeper ground–based images in which we believe we have identified the likely absorber.

## 2. Observations and Reductions

Optical spectra of PKS 0454+0356 were obtained on the night of 24 December 1992 using the Lick Observatory 3m Shane telescope and the Kast double spectrograph; the blue–side spectra were obtained with an 830 line/mm grism blazed at 3460 Å and providing a spectral resolution of 2.3 Å (2 pixels) over the wavelength range 3280–4620 Å. The total integration time was 7200s. Relevant portions of the blue spectrum are plotted in Figure 1. A summary of absorption line measurements relevant to the abundances in the $z_{abs} = 0.8596$ system is given in Table 1.

Optical images of the PKS 0454+0356 were obtained as part of a large survey for galaxies producing Mg II absorption line systems (Steidel *et al.* 1995a,b), including a very deep $\mathcal{R}[6930/1500]$ image obtained in 1994 January at the 2.4m Hiltner telescope of the Michigan–Dartmouth–MIT Observatory. This image consisted of a total integration of 7600s and the final co-added image has FWHM=0″.89. A deep $K$ band image was obtained in 1993 March using the 256×256 NICMOS 3 IRIM on the Kitt Peak 4m Mayall telescope. All of the images were reduced using standard techniques, described in detail elsewhere (Steidel *et al.* 1995b).

The *HST Faint Object Spectrograph* (FOS) was used with the G270H grating and red digicon on 1994 July 26 to obtain an ultraviolet spectrum of PKS 0454+0356. The expected DLA line at $z_{abs} = 0.8596$ is at the blue end of the observed spectral range.

## 3. The H I Column Density at $z_{abs} = 0.8596$

The pipeline-calibrated HST spectrum was resampled to the original dispersion of 0.51 Å pix$^{-1}$. A small constant flux ($\sim 3\%$ of the continuum) was subtracted from the data to place the base of the DLA line with exactly zero flux. At the position of the DLA line, the spectrum could not be normalized using data points on either side of the line because of a lack of continuum to the blue of the absorption, and broad O VI emission lines from the QSO to the red. Regions free of emission were defined over the whole observed spectrum and a power law fitted. The spectrum was normalized using this fit, and theoretical damped Ly$\alpha$ line profiles were then fitted to the data. Details of the routines used to derive $N(\text{H I})$ and errors associated with Poisson statistics are given in Bowen, Blades & Pettini (1995). The best fit of $N(\text{H I}) = 5.7 \times 10^{20}$ cm$^{-2}$ is shown in Fig. 2. The uncertainty in $N(\text{H I})$ is dominated by the error in the continuum placement, and is estimated to be $\leq 0.3 \times 10^{20}$ cm$^{-2}$ by simply using various plausible power laws and refitting the line. Line profiles of $N(\text{H I})$ different by $\pm 2\sigma$ of the best fit are also shown in Fig. 2.

Using the measured H I column density for this system, we can set a lower limit on the spin temperature in the H I gas using the upper limit of $\tau_{21} < 0.010$ set by Briggs & Wolfe (1983) in their survey for 21–cm absorption associated with Mg II–selected absorption line systems. Using the values obtained by them, and the H I column density measured by us, we find $T_s > 1740 \left( \frac{b}{10 \text{ km s}^{-1}} \right)^{-1}$ K , where $b$ is the Doppler parameter associated with the absorbing gas. A conservative lower limit on $T_s$ is obtained by assuming the $b$ value of 34 km s$^{-1}$ obtained from a single component curve of growth analysis (see §4 below) of the Fe II lines, $Ts > 580$K. Thus, $T_s$ is much higher than is typically found in dense H I clouds in the Galactic disk, but is in line with other limits set for high redshift damped Lyman $\alpha$ systems (see, e.g., Briggs & Wolfe 1983, Briggs 1988).

## 4. The Zn, Cr, and Fe Abundances

Our resolution is insufficient to resolve the Zn II/Mg I blend at $\lambda 2026$ and the Cr II/ZnII blend at $\lambda 2062$. However, we have a measurement of the Mg I $\lambda 2852$ line at $z_{abs} = 0.8596$ from SS92. By assuming that the Mg I lines are on the linear portion of the curve of growth, the *expected* strength of Mg I $\lambda 2026$ is only 12mÅ . In addition, the identification of the line at $\lambda 3767.81$ with Zn II leads to a much better agreement with the systemic redshift of $z_{abs} = 0.85965 \pm 0.00005$ determined from all of the well–detected low–ionization lines in the system. We therefore assign a strength of 41 mÅ for the Zn II $\lambda 2026$ line. If we then assume that the Zn II $\lambda 2062$ line is half the strength of $\lambda 2026$ (i.e., linear curve of growth), then the Cr II $\lambda 2062$ strength is 74mÅ, so that the ratio of the $\lambda 2062$ to the $\lambda 2056$ lines agrees very well with the relative $f$ values. Cr II $\lambda 2066$ is not reliably detected, but the limit on its strength also supports the assertion that the Cr II lines are in the ratio expected for unsaturated lines. We have measurements for a total of 7 Fe II lines representing a wide range of $f$ values, and a curve of growth analysis gives log N(Fe II) = $15.02 \pm 0.07$ and an effective $b = 34 \pm 2$ km s$^{-1}$. Our assertion that saturation effects are unimportant for the weak Zn II and Cr II lines is consistent with the findings of Pettini *et al.* (1994), and also with the handful of very high resolution observations of higher redshift Zn II and Cr II lines (Wolfe *et al.* 1994); however, we cannot *prove* that this is the case without much higher resolution data. In Table 2 we summarize the inferred column densities of Zn II, Cr II, and Fe II, and also the gas–phase abundances of each element assuming that $N(X^+)/N(H\ I) = N(X)/N(H)$.

The overall metallicity (as reflected by [Zn/H]) in the gas at $z_{abs} = 0.8596$ is quite similar to a "typical" damped Lyman $\alpha$ system at much higher ($z \sim 2$) redshift (Pettini *et al.* 1994). We also see the same trend for the depletion of Cr II relative to Zn II to be much less important than in the local ISM of the Galaxy. In fact, compared to all of the damped Lyman $\alpha$ systems observed by Pettini *et al.* (1994) for which actual measurements of both N(Zn II) and N(Cr II) were made, we find that the PKS 0454+0356 system has the among the *highest* ratio N(Cr II)/N(Zn II)[5],

---

[5]We note that Pettini *et al.* 1994 used a Zn II $f$ value that is smaller by $\sim 20\%$, so that for direct comparison

suggesting that depletion onto dust grains may be even less important in the ISM of the $z = 0.8596$ galaxy than in many of the higher redshift counterparts. The measurements of [Cr/H] and [Fe/H] relative to [Zn/H] are actually consistent with *no* depletion of Cr and Fe onto dust grains, as compared to the 98%–99% of those elements inferred to be in solid form in the local ISM. Thus, the $z_{abs} = 0.8596$ damped Lyman $\alpha$ system has comparable metal abundances, but perhaps a significantly lower dust/gas ratio for its metallicity, as compared to the high redshift systems.

## 5. The Galaxy Identification

In Figure 3 we show a contour plot of the deep $\mathcal{R}$ image after the QSO light profile has been subtracted by modeling the point spread function (PSF) using bright stars in the field. We have identified the $\mathcal{R} = 24.6$ galaxy $2''\!.1$ N of the QSO as the most likely candidate absorber, based on the absence of anything brighter within a projected distance of $\sim 20h^{-1}$ kpc of the QSO sightline which corresponds to the likely range expected for such a high Mg II equivalent width system (of a sub–sample of 16 Mg II absorbers having Mg II rest equivalent widths $> 1.5$Å, only one has an impact parameter $> 20$ kpc [Steidel *et al.* 1995b]). We could have detected an object of comparable brightness to G3 at any angular separation $> 0''\!.8$ from the QSO sightline. If placed at $z = 0.859$, the centroid of G3 is $8.7(10.8)h^{-1}$ kpc from the QSO line of sight for $q_0 = 0.5\ (0.05)$, which is within the range typically observed for similarly strong Mg II systems . We can only set limits on the $K$ magnitude of the object, $K > 20.5$. The corresponding limit $\mathcal{R} - K < 4.1$ makes the object bluer than an unevolved Sa galaxy at $z = 0.859$, so that it is not particularly constraining. There are no galaxies detected in our $K$ band image to $K \sim 20.5$ that are both within $4''$ of the QSO and undetected in the optical passband. Adopting a k–correction appropriate for a late–type spiral galaxy at $z = 0.859$, galaxy G3 would have an absolute $B$ magnitude of $M_B = (-19.5, -19.0)$ for $q_0 = (0.05, 0.5)$, placing it within the lowest 10% in luminosity of the sample of 58 Mg II absorbers identified by Steidel *et al.* (1995b). There is an additional known Mg II absorption system at $z_{abs} = 1.1537$ (Steidel & Sargent 1992); this latter system is much weaker in terms of the Mg II absorption line strength and the corresponding Lyman $\alpha$ line is clearly *not* damped on the basis of the *FOS* spectrum. If the galaxy indicated were to be identified with the $z_{abs} = 1.1537$ system instead, it would mean, of course, that the $z_{abs} = 0.8596$ absorber must be even *fainter* than the suggested object.

## 6. Discussion

The low abundances in the $z_{abs} = 0.8596$ DLA system present another puzzling challenge to the idea that, if the damped systems are tracing the progenitors of disk galaxies, then at look–back

---

their derived column density must be divided by $\sim 1.2$.

times corresponding to $z \sim 0.85$ they should already have developed near–solar metal abundances. As summarized in Table 2, *both* examples of $z < 1$ DLAs for which abundance determinations have been made are strikingly similar in their abundance and depletion to the high redshift systems. On the other hand, direct imaging of the galaxies in both cases have suggested that the galaxies responsible for the absorption are in some way "unusual", with the 3C 286 galaxy being of low surface brightness and the present case being a somewhat underluminous galaxy in comparison to, e.g., the Milky Way. In view of the galaxy properties in both of these cases, it is not then all that surprising that the abundances and dust content are well below the "expectation" for galaxies on an evolutionary track similar to that followed by the Milky Way. Again, more observations of moderate redshift DLAs are needed to establish any significant trends; however, we note in passing that there are several known 21-cm absorbers and DLAs in the Mg II absorbing galaxy sample of Steidel *et al.* (1995b), and the distribution in their luminosities and colors is rather different from the general distribution for the absorbing galaxies, in the sense that they tend to be (on average) less luminous and bluer. Certainly none of the galaxies is similar in its overall color and luminosity to the Milky Way; however, there are many galaxies in the Mg II absorber sample that *do* resemble Milky–Way like systems, and we do not have complete knowledge of the associated H I column densities. Nevertheless, there is already the suggestion that there may be a selection effect against luminous spiral galaxies for moderate redshift DLA systems, perhaps stemming from the difficulty in actually discovering QSOs (optically) that are intersecting chemically evolved galaxies with Milky–Way–like dust content. Only more observations, particularly using *HST* to measure H I column densities, can confirm or refute this suggestion.

Analysis of HST data by DVB and JCB was funded from NASA grant GO-3525.01-91A. We thank Ken Sembach for help with the curve of growth analysis, and Max Pettini for valuable comments on an earlier draft.

Table 1.   Zn II, Cr II, and Fe II Absorption Lines[a]

| Line No. | $\lambda_{obs}$ (Å) | $\sigma(\lambda)$ (Å) | $W_0$ (mÅ) | $f$[b] | ID | $z_{abs}$ |
|---|---|---|---|---|---|---|
| 1 | 3767.81 | 0.32 | 53 ± 12 | 0.515 | ZnII(2026.14) | 0.8596 |
| | | | | 0.115 | MgI(2026.48) | 0.8593 |
| 2 | 3823.96 | 0.21 | 102 ± 12 | 0.140 | CrII(2056.25) | 0.8597 |
| 3 | 3834.93 | 0.23 | 94 ± 13 | 0.105 | CrII(2062.23) | 0.8596 |
| | | | | 0.253 | ZnII(2062.66) | 0.8592 |
| 4[c] | 3841.69 | 0.43 | 34 ± 13 | 0.070 | CrII(2066.16) | 0.8593 |
| 5 | 4183.92 | 0.19 | 101 ± 12 | 0.0025[d] | FeII(2249.88) | 0.8596 |
| 6 | 4204.30 | 0.13 | 110 ± 10 | 0.0037[d] | FeII(2260.78) | 0.8597 |

[a]Features indicated in Figure 1 only

[b]Oscillator strength, from Morton (1991).

[c]This line is not considered to be a signficant detection.

[d]These $f$ values are quite uncertain; a full analysis using 7 Fe II lines suggests that both these lines should be on the linear part of the curve of growth (see §4 of the text).

Table 2.   Properties of $z < 1$ Damped Lyman $\alpha$ Systems[a]

| | PKS 0454+0356[b] | 3C 286[c] |
|---|---|---|
| $z_{abs}$ | 0.8596 | 0.6920 |
| log N(HI) | $20.75 \pm 0.03$ | $21.28 \pm 0.03$ |
| log N(Zn II) | $12.34 \pm 0.08$ | 12.70 |
| log N(Cr II) | $13.30 \pm 0.04$ | 13.18 |
| log N(Fe II) | $15.02 \pm 0.08$ | 14.95 |
| [Zn/H] | $-1.1$ | $-1.2$ |
| [Cr/H] | $-1.2$ | $-1.8$ |
| [Fe/H] | $-1.2$ | $-1.8$ |

[a]All column densities are in units of $cm^{-2}$, and the abundance measurements are logarithmic abundances relative to the solar values tabulated by Anders & Grevesse (1989)

[b]This paper.

[c]Meyer & York 1992

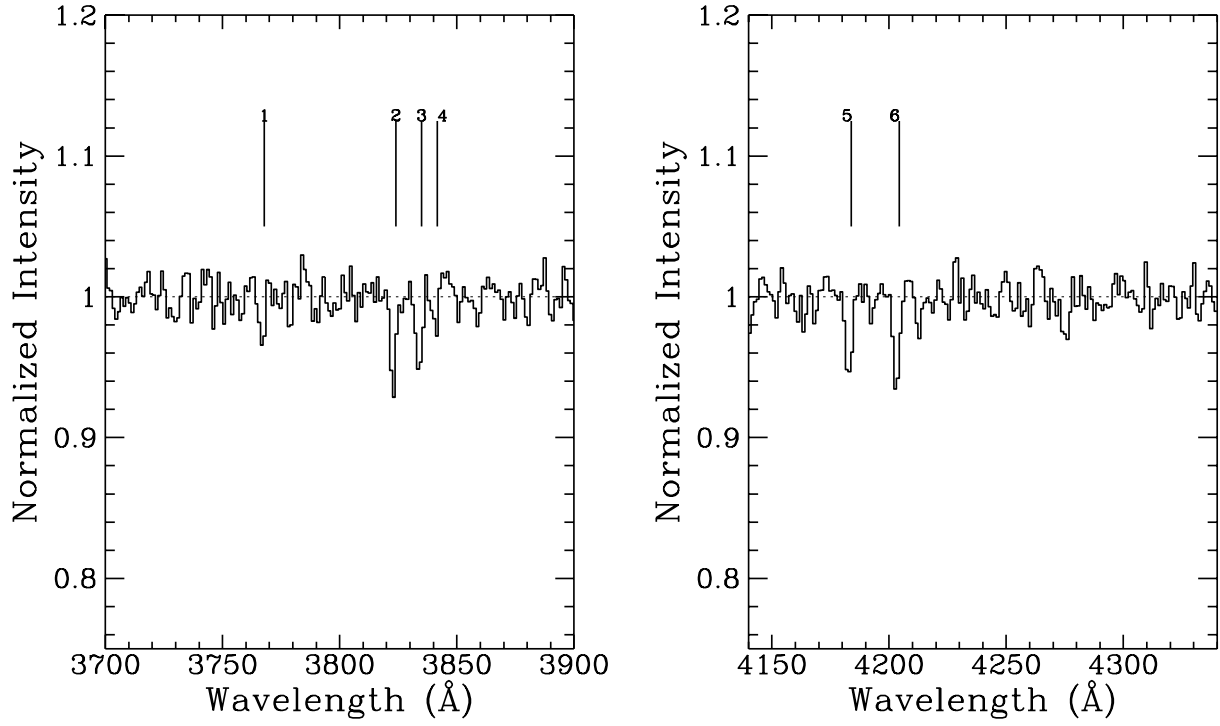

Fig. 1.— Portions of the Lick spectrum of PKS 0454+0356, showing the Zn II and Cr II (left panel) and the Fe II $\lambda 2248$ and $\lambda 2260$ lines (right panel) from the $z_{abs} = 0.8596$ system (see Table 1).

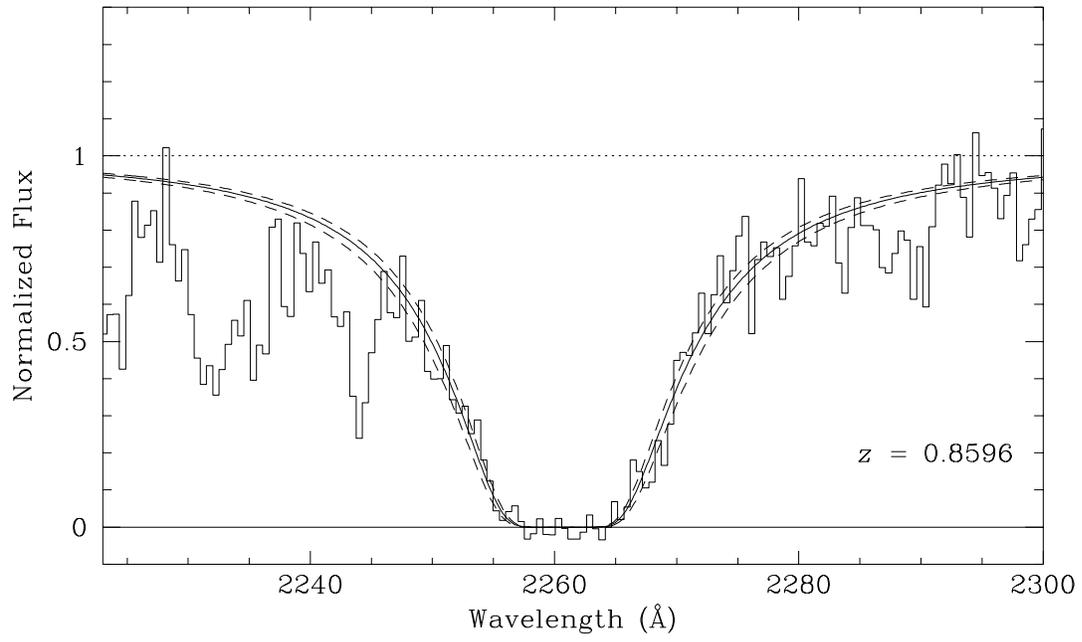

Fig. 2.— Plot showing the damped Lyman $\alpha$ profile, and the fits which gives N(H I)= $5.7\pm0.3\times10^{20}$ cm$^{-2}$. The dashed curves represent $\pm2\sigma$ from the best–fit value.

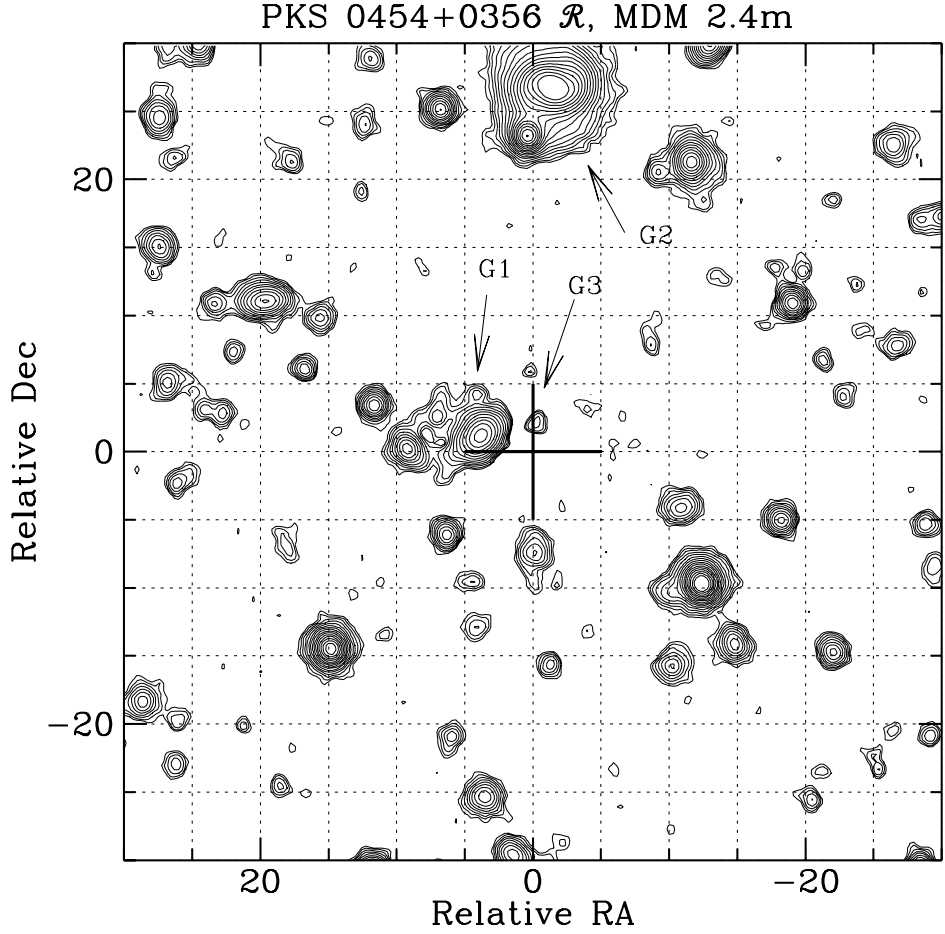

Fig. 3.— A contour plot of the deep $\mathcal{R}$ band image of the PKS 0454+0356 field, where the QSO has been subtracted to reveal underlying objects. Galaxy G1 and G2 are at $z = 0.072$ and $z = 0.201$, respectively (Steidel *et al.* 1993), and G3 is our tentative identification of the $z_{abs} = 0.8596$ absorbing galaxy.